\documentclass[12pt, draftclsnofoot, onecolumn]{IEEEtran}

\usepackage{amsmath, amssymb, amsfonts, amssymb, amsthm}
\usepackage{bbm}
\usepackage{graphicx} \graphicspath{ {figure/} }
\usepackage{epstopdf} 
\usepackage[noabbrev]{cleveref}
\usepackage{scalefnt}
\usepackage{indentfirst}
\usepackage{cite}
\usepackage{bm}
\usepackage{enumerate}
\usepackage{enumitem}
\usepackage[caption=false,subrefformat=parens,labelformat=parens]{subfig}
\usepackage{scalerel}
\usepackage{scalerel}
\usepackage{algpseudocode}
\usepackage{multirow}
\usepackage[table,xcdraw]{xcolor}

\usepackage{algorithm}
\usepackage{algcompatible}
\renewcommand{\algorithmiccomment}[1]{\hfill ~#1}
\algnewcommand\INPUT{\item[\textbf{Input:}]}
\algnewcommand\INITIAL{\item[\textbf{Initialization:}]}
\algnewcommand\OUTPUT{\item[\textbf{Output:}]}
\algnewcommand\RETURN{\item[\textbf{Return:}]}
\algnewcommand\ITER{\item[\textbf{Iteration:}]}
\algrenewcommand\algorithmiccomment[2][\small]{{#1\hfill\ #2}}

\theoremstyle{plain}

\renewcommand{\algorithmiccomment}[2][.5\linewidth]{\leavevmode\hfill\makebox[#1][l]{//~#2}}

\theoremstyle{definition}

\theoremstyle{remark}

\begin{document}
\title{Towards Intelligent Millimeter and Terahertz Communication for 6G: Computer Vision-aided Beamforming}

\author{Yongjun Ahn, Jinhong Kim, Seungnyun Kim, Kyuhong Shim, Jiyoung Kim, Sangtae Kim, and Byonghyo Shim}

\maketitle

\begin{abstract}
Beamforming technique realized by the multiple-input-multiple-output (MIMO) antenna arrays has been widely used to compensate for the severe path loss in the millimeter wave (mmWave) bands.
In 5G NR system, the beam sweeping and beam refinement are employed to find out the best beam codeword aligned to the mobile.
Due to the complicated handshaking and finite resolution of the codebook, today's 5G-based beam management strategy is ineffective in various scenarios in terms of the data rate, energy consumption, and also processing latency.
An aim of this article is to introduce a new type of beam management framework based on the computer vision (CV) technique.
In this framework referred to as computer vision-aided beam management (CVBM), a camera attached to the BS captures the image and then the deep learning-based object detector identifies the 3D location of the mobile.
Since the base station can directly set the beam direction without codebook quantization and feedback delay, CVBM achieves the significant beamforming gain and latency reduction.
Using the specially designed dataset called Vision Objects for Beam Management (VOBEM), we demonstrate that CVBM achieves more than 40$\%$ improvement in the beamforming gain and 40$\%$ reduction in the beam training overhead over the 5G NR beam management.
\end{abstract}

\newpage

\section{Introduction}
As an answer to the ever-increasing demands on the data rate, reliability, latency, and connectivity, wireless industry began to explore the high frequency bands in the spectrum above $24\,$GHz that have long been considered unsuitable for the wireless communication.
When compared to the systems using microwave, the main bottleneck of this so-called mmWave band is the short communication distance caused by the high diffraction and penetration loss, atmospheric absorption, and rain attenuation.
To compensate for the severe path loss, beamforming techniques realized by the multiple-input-multiple-output (MIMO) antenna arrays have been widely used.
Since the beamforming gain is maximized only when the beams are properly aligned with the signal propagation paths, the base station (BS) needs to acquire the accurate downlink channel information in a form of angle of arrival/departure (AoA/AoD) and distance in the beam generation.

Beam management process in 5G NR to support the channel information acquisition and beam generation consists of two steps.
In the first step, called the \textit{beam sweeping} stage, the BS transmits the set of synchronization signal block (SSB) beam codewords, each of which covers the relatively wide physical area.
After receiving SSB, the UE measures the reference signal received power (RSRP) of the SSB beams and then feeds back the index of the SSB beam corresponding to the largest RSRP.
In the second step, called the \textit{beam refinement} stage, the BS identifies the mobile user's location by sending multiple pilot signals to the direction determined by the beam sweeping stage.

While the beam management process in 5G NR is straightforward and intuitive, it requires a complicated and laborious handshaking process to identify the optimal beam direction between the BS and mobile device.
In fact, due to the dual closed-loop architecture of the beam management process, a significant beam training overhead in terms of power, latency, and resource management is unavoidable.
Thus, even with a small movement of the mobile, say in the order of a few meters, the beam goes awry, resulting in a re-start of the beam training process.
Since a simple remedy such as the frequent transmission of beams (beam sweeping period of current 5G NR is $20\,$ms~\cite{TR38802}) will waste resources, increase the latency, and also requires higher power consumption, beam management mechanism in 5G and its simple extension might not be suitable for the 6G era where more high frequency terahertz (THz) band is employed.

The primary goal of this article is to introduce a computer vision (CV)-aided framework to control the beam management in mmWave and THz communication regime.
In contrast to the traditional 5G-based beam management (5G-BM) relying exclusively on the radio frequency (RF) transmission of control/pilot signals, we exploit the CV technique in identifying the beam transmit direction.
Our approach is justified by two crucial observations that 1) physical characteristics of mmWave and THz radio waves are getting closer to the visible light (400$\sim$790 THz) in that the transmit energy is mostly concentrated in the line of sight (LoS) path and 2) recent advances in deep learning (DL)-based CV techniques have made a gigantic leap in performing the object classification, detection, and tracking from raw images.
Using the images obtained from the imaging sensors such as RGB camera, LiDAR, radar, or combination of these, DL-based object detector extracts the geometric information (azimuth/elevation angle and distance of a mobile) using which the transmit beam can be generated without quantization and feedback.

The main contributions of this paper, particularly when compared to the conventional approaches based on the codeword-based beamforming~\cite{ref_new1, ref_new2, ref_new3, ref_new4}, are summarized as follows:

\begin{itemize}
\item We propose a new framework referred to as the CV-aided beam management (CVBM) that exploits the CV technique in identifying the mobile user's location.
In CVBM, using the image obtained from the RGB-depth (RGB-d) camera, the DL-based object detector identifies the location of a mobile, meaning that the BS can transmit the information-bearing beam without the codebook quantization and handshaking process. 
In doing so, the quality of beamforming can be greatly enhanced and the transmit power and the control plane latency can be reduced over the conventional 5G-BM.

\item We generate a special dataset referred to as the \textit{Vision Objects for Beam Management} (VOBEM) to evaluate the effectiveness of the proposed framework.
The VOBEM dataset consists of 135 pairs of RGB and depth images labeled with categories (person or mobile) and bounding boxes of the objects.
In contrast to the well-known image datasets such as MS-COCO 2017 dataset~\cite{mscoco} and ImageNet dataset~\cite{imagenet}, VOBEM includes the samples obtained from the viewpoint of the RGB-d camera attached to the BS.

\item We develop the two-stage DL-based 3D localization scheme for better detection of mobile device in an image.
In the first stage, we perform the \textit{large-scale object detection} to identify a person holding a mobile device.
We then perform the \textit{small-scale object detection} to identify the mobile device (see Section III. A for details).

\item We test the localization, latency, energy consumption, and data rate performances from the location information inferred from the DL-based object detector.
From the numerical evaluations on the downlink transmission scenario of (sub-)THz communication, we demonstrate that CVBM achieves more than $40\%$ improvement in the beamforming gain and $40\%$ reduction in the beam training overhead over the 5G-BM.
\end{itemize}

The rest of this article is organized as follows.
In Section II, we briefly review the conventional beam management and CV techniques.
In Section III, we explain the principle of CVBM and the object detection operation to extract 3D location of a mobile.
In Section IV, we present the simulation results of the THz-band beamforming.
We discuss future issues and conclude the paper in Section V.

\section{Basics of Conventional Beam Management and Computer Vision Paradigm}
In this section, we provide a brief overview of the beam management process of 5G NR and then discuss the DL-based object detection technique in CVBM.

\subsection{Beam Management in Conventional 5G Communication Systems}
In the beam management of 5G NR, a beam codebook, a set of analog beam codewords (pre-defined analog phase shift values generated from the DFT matrix), is used~\cite{mmWave_dd}.
To find out the best beam codeword aligned to the mobile, two-step processes called the beam sweeping and beam refinement have been introduced.
In the beam sweeping step, the BS transmits a sequence of SSB beams chosen from the beam codebook to find out the rough estimate of the user location (e.g., azimuth/elevation angles in the intervals of 45 degrees).
Among these, a mobile picks the beam codeword maximizing the RSRP and then feeds back the index of chosen beam codeword to the BS.
In the beam refinement step, the BS narrows down the choice of the beam by transmitting multiple channel state information-reference signal (CSI-RS) beams to the direction designated by the beam sweeping process.
Once the beam codeword is chosen, the BS transmits the data to the direction designated by the beam codeword.
In summary, the detailed beam management process in 5G NR is as follows (see Fig.~\ref{fig_5G_beam_management}):

\begin{figure*}[t]
	\centering
	\includegraphics[width=1\columnwidth]{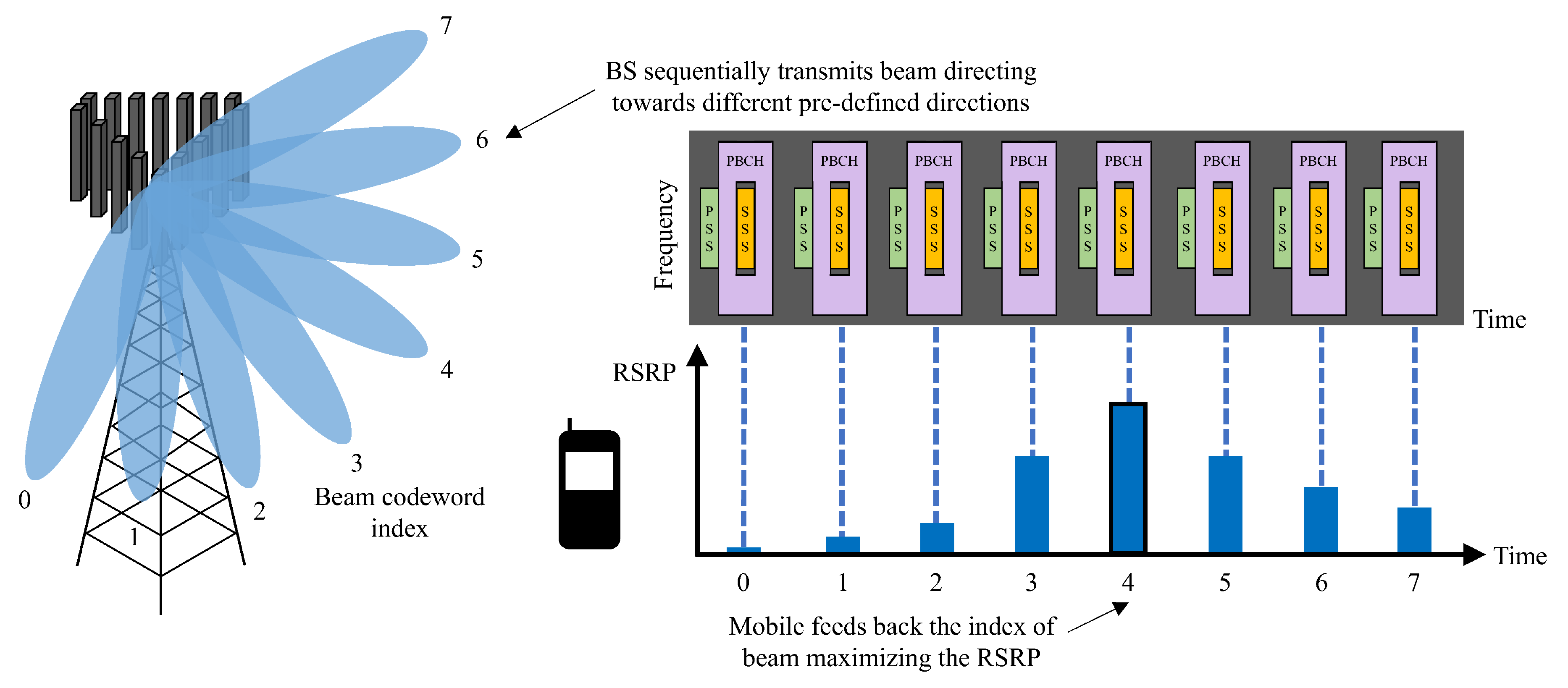}
	\caption{Illustration of beam management in 5G NR. SSB consisting of the primary synchronization block (PSS) and secondary synchronization block (SSS) is transmitted via physical broadcast channel (PBCH). Note that SSS and PSS are used to achieve the frame and symbol-level synchronization between the BS and UE, respectively. Basically, complicated handshaking operation based on beam sweeping and measurement feedback is required to find out the optimal pair of transmit and receive beams.}
	\label{fig_5G_beam_management}
\end{figure*}

\begin{enumerate}
\item In the first step, the BS transmits multiple beams, each of which is directing toward a pre-defined direction. In this operation called SSB beam sweeping, the beam carries the SSB to broadcast the system information and synchronize the mobile with the BS. 
Typically, $4$ and $8$ SSB beams (this set of SSB beams is referred to as the SSB burst set) are transmitted within $5\,$ms but the number of SSB beams can be increased up to $64$ in the mmWave band in $28\,$GHz.
After the SSB beam sweeping, the mobile feeds back the index of the best SSB beam to the BS.
\item After the initial beam-pair link establishment, to narrow down the beam transmission direction, the BS transmits beams carrying the CSI-RS within the physical range designated by the SSB beam.
In this so called beam refinement step, up to $4$ CSI-RS beams are transmitted within $30\,$ms.
After receiving these, a mobile measures the RSRP and then feeds back the index of the CSI-RS beam maximizing the RSRP to the BS using which the BS determines the transmit beam direction.
\end{enumerate}

Since the beam alignment should be done with limited time/frequency resources and transmit power, codebook-based beamforming scheme has been employed.
Main problem of this approach is that the angle mismatch between the pre-defined beam direction and the real direction will cause a significant degradation of the beamforming gain.
For example, in the 5G NR system with an $8\times 8$ planar antenna array, up to $256$ CSI-RS beams can be used.
In this case, each beam covers the circle sector of $15.8$ square degrees on average so that the beam direction error in the worst case will be $7.9$ degrees in azimuth and elevation angles.
Since the half power beam width, the range of angle where the relative beamforming gain is more than 50$\%$ of the peak gain of beam, of the system operating in $28\,\text{GHz}$ FR2 band is around $10$ degrees~\cite{mmWave_beamwidth}, the beam direction mismatch will degrade the beamforming gain, ending up with the considerable rate loss and the power waste.
This issue will be more pronounced in the 6G systems employing THz band since the THz beam width is much narrower than the mmWave beam width (e.g., around 3 degrees for $32\times 32$ antenna system in 0.3 THz band~\cite{THz_beamwidth}).

Another serious problem is the latency caused by the complicated handshaking process between the BS and mobile.
In the beam refinement operation of 5G NR systems, each CSI-RS beam is transmitted with $10\,\text{ms}$ period so that it takes around $30\,\text{ms}$ to transmit $4$ CSI-RS beams. 
Since the coherence time of mmWave/THz channel is much shorter than that of microwave channel (e.g., $9\,\text{ms}$ when the mobile is moving at $30\,\text{km/h}$ speed), the beam direction will be totally misaligned even with a small movement of the mobile.
In this case, the beam management process should be started over, increasing the latency for the beam establishment.
For these reasons mentioned above, many network operators deploy the mmWave solutions only for limited scenarios (e.g., fixed wireless access).
Recently, there have been many studies considering an improvement of the codebook design, reduction of the angle quantization, and proposal of new antenna array structure as well as the beam control period reduction~\cite{ref_new1, ref_new2}.
While these approaches might be effective to some extent, they will impose fundamental limits on beamforming accuracy and beam management latency due to the humongous number of codewords, extra hardware, and complicated handshaking process.

\subsection{Computer Vision and Object Detection}
\begin{figure}
	\centering
	\subfloat[]{\includegraphics[width=0.96\columnwidth]{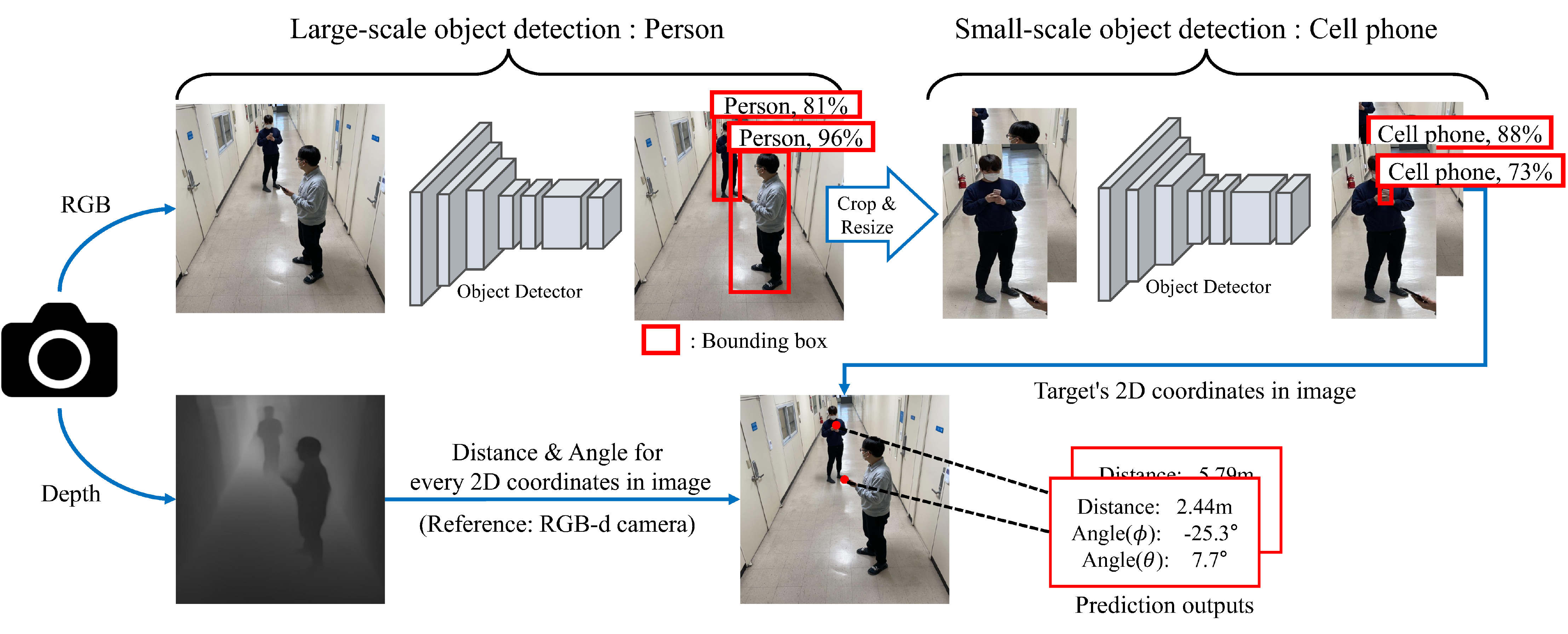}}
	\hfill
	\subfloat[]{\includegraphics[width=0.8\columnwidth]{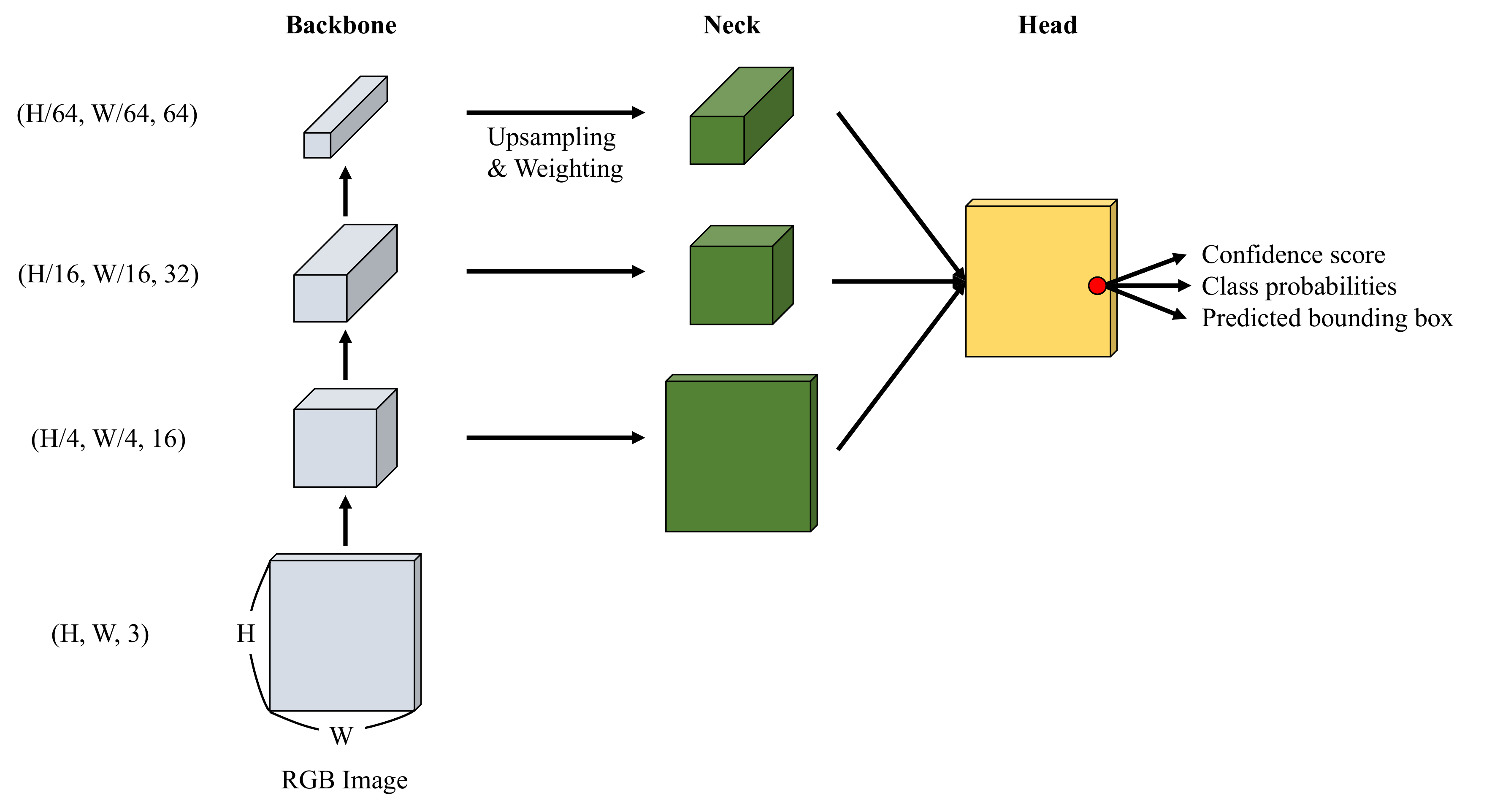}}
	\hfill
	\caption{(a) Illustration of vision information extraction process. DL-based object detector is used to find out the target object location in the 2D image (top row). Then, the prediction result is integrated with the depth information to generate 3D coordinates (bottom row). The percentage indicates the probability of the detected object belonging to the specific class (i.e., class score). (b) The detailed architecture of the DL-based object detector including three sub-structures; backbone, neck, and head. (H, ,W, 3) denotes the size of the RGB image in terms of height, width, and red/green/blue color maps.}
	\label{fig:vision_system}
\end{figure}

CV is a field that aims to extract useful information from images and videos.
Due to the recent advances in the DL model and architecture, DL-based CV has surpassed the performance of humans in many applications including face detection, medical image analysis, visual surveillance, to name just a few~\cite{face_det}.
DL-based CV is distinct from the conventional CV algorithms in that the task is trained with DNN using the abundant and properly designed dataset.
In the training phase, DL parameters (weights and biases) are updated to identify the end-to-end mapping between the input images and the desired outputs.

In CVBM, we primarily use the DL-based object detection to identify the beam direction.
The main goal of the object detection is to find out the position (2D cartesian coordinate $(x, y)$) of target objects in an image and then determine its class (e.g., human, mobile, or vehicle).
By identifying the target object such as the mobile phone, antenna, or laptop and then measuring the angle and distance from the BS to the object, we can accurately identify the beam direction with negligible processing (DL-inference) delay.
Specifically, the first step of the DL-based object detector is to find out the class (in our case, person or mobile) and the bounding box, the smallest rectangular-shaped box containing the target object (see Fig.~\ref{fig:vision_system}(a)).
Since the bounding box contains the upper-left and the lower-right cartesian coordinates of the selected rectangular area, we can easily compute the centroid of the object by averaging two coordinates.

Since the input image has a spatial structure, convolutional neural network (CNN) is one natural option for the DNN architecture.
Due to the local connectivity of the convolution filter kernel, CNN facilitates the extraction of spatial correlated feature from the raw image.
To scrutinize the quality of the CNN-based object detection model and then reflect it in the weight update process, we employ three different training losses for each pixel: 1) cross entropy measuring the difference between the ground-truth probability of a pixel (1 if a pixel belongs to the object and 0 otherwise) and the estimated probability called confidence score, 2) mean squared error (MSE) evaluating the position error of the bounding box in terms of width, height, and center position of object, and 3) another cross entropy measuring the class prediction error of a pixel for two classes (i.e., person and mobile).

\section{Vision-Aided Beam Management for 6G mmWave and THz MIMO Systems}
In this section, we discuss the overall process of CVBM and the vision information extraction using the DL-based object detection.

\subsection{Vision Information Extraction Process}
Before performing the 3D localization of a mobile device, we need to process the DL-based object detection on the 2D image.
For the object detection, we exploit the CNN consisting of three main components: \textit{backbone}, \textit{neck}, and \textit{head} (see Fig.~\ref{fig:vision_system}(b)).
The backbone extracts the features (e.g., color, shape, and face) in an image.
Then, the neck puts different weights on the local features (e.g., edge and curve) and global features (e.g., face and wall) so that the head can focus only on the information needed for the object detection and class prediction.
Using the weighted features as inputs, the head computes the confidence score, class score, and height and width of the bounding box for each pixel on the image.
Confidence score indicates the likelihood of the point being the center of the object, so that the point with the highest confidence score is determined to be a center.
The class score measures the probability of each object belonging to class (i.e., human or mobile phone).
Using the confidence score and the class score, the DL-based object detector generates the class and 2D position of the bounding box.

Since the 3D location of the target object (in our example, a mobile) is needed for the beamforming, we additionally should have the distance information between the BS and object.
To this end, we use a RGB-d camera, an imaging device equipped with the RGB camera and LiDAR sensor. 
A notable feature of RGB-d camera is that it can measure the distance to the point in each pixel using the LiDAR sensor.
Using the depth information $r$ provided from the RGB-d camera together with the 2D coordinates $(x,y)$ of the target object in the image, the 2D position can be converted into the 3D cartesian coordinates $(x,y,z)$ (i.e., $z = \sqrt{r^2-(x^2 + y^2)}$).
Then the spherical coordinate $(r,\theta,\phi)$ is obtained using the equations $\theta=\arctan\frac{\sqrt{x^2 + y^2}}{z}$ and $\phi=\arctan{\frac{y}{x}}$.

We would like to mention that the performance of the DL-based object detector is sensitive to the distance.
This is because when the camera is far away from the mobile, only a few pixels represent the mobile in the captured image.
To handle the problem, we design the two-stage DL-based 3D localization process.
In the first stage, we perform the \textit{large-scale object detection} to find out a person holding a mobile device.
To find out the bounding box of a person, the whole image is used as an input of the DL-based object detector.
Then the \textit{small-scale object detection} is performed to identify the mobile device.
In this stage, using the small-sized bounding box containing a person as an input (see Fig.~\ref{fig:vision_system}(a)), the DL-based object detector generates the 3D location of a mobile.

\subsection{Overall Process}
\begin{figure*}[t]
	\centering
	\includegraphics[width=1\columnwidth]{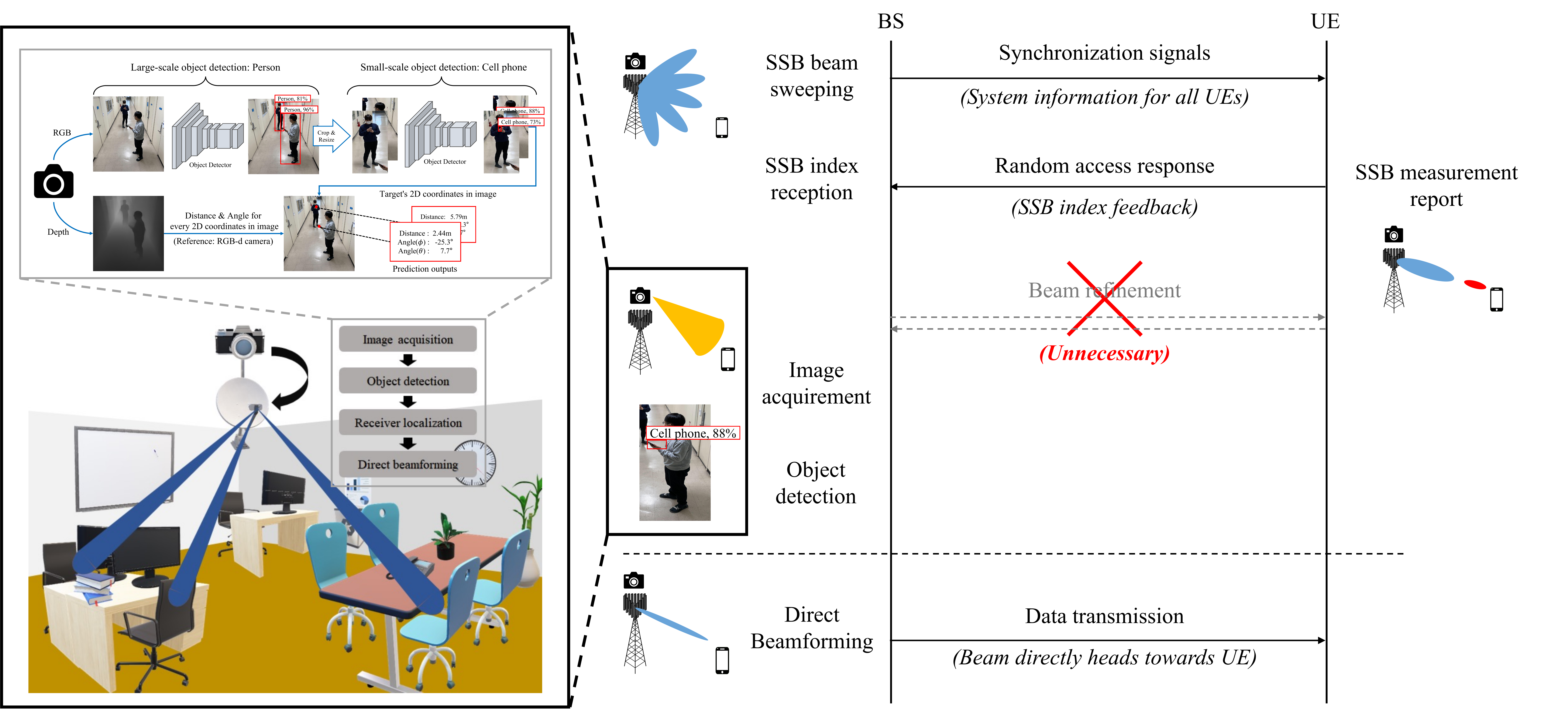}
	\vspace{-0.2cm}
	\caption{Illustration of CVBM. The beam refinement step is replaced by the image acquisition and object detection.}
	\label{fig:vision_aided_beam}
\end{figure*}

The holy grail of CVBM is to extract the location information of a mobile using the object detection and then directly use this information in the beam generation.
Among several implementation options, we here consider a hybrid beam management strategy that exploits the conventional SSB beam sweeping in the initial access stage and then determines the final beam direction using the object detection (see the illustration in Fig.~\ref{fig:vision_aided_beam}).
In this scheme, after the SSB beam sweeping, the BS captures the image from the camera within the physical area designated by the SSB beam index.
Once the azimuth and elevation angles of a mobile are identified via the object detection, by controlling the analog phase shifts of antenna elements, the BS generates the directional beam.
For example, when the planar antenna array is used, the beam is generated by the Kronecker product of azimuth and elevation array steering vectors.
The detailed operations of CVBM are as follows:
\begin{itemize}
\item[1)] The BS performs the initial access by receiving the feedback on SSBs from the mobile.
In this beam sweeping stage, each SSB beam covers a relatively wide angle.
For example, when the BS uses $8\times 4$ SSB beams, each beam covers azimuth/elevation angles in the intervals of 45 degrees.
\item[2)] The RGB-d camera takes a shot for the area covered by the SSB beam index. 
The image can be captured from the camera periodically, aperiodically, or semi-persistently. 
From the obtained image, DL-based object detector finds out the location of a mobile in the form of 3D spherical coordinate $(r_{m},\theta_{m},\phi_{m})$ where the location of the BS is used as the reference point $(r_{b}=0,\theta_{b}=0,\phi_{b}=0)$.
Using the azimuth and elevation angle estimates $(\theta_{m},\phi_{m})$, the BS generates the narrow directional beam such that the main lobe of a beam radiation pattern is directed toward the mobile's location.
The BS exploits the estimated distance $r_m$ between the BS and mobile to control the beam power.
\end{itemize}

When compared to 5G-BM, CVBM has a number of benefits.
\begin{itemize}
\item \textit{Beamforming gain maximization}: In contrast to the codebook-based beam management where the mismatch between the pre-defined discrete beam direction and the real beam direction is unavoidable, the proposed vision-aided beam management is free from discretization error since the beam is heading towards the location of the target mobile.
Through extensive simulations, we numerically confirm that the direction error of the DL-based object detection is less than $0.5$ degrees (see Table I).
Considering the rapid development of the CMOS sensor technology, we will have better pixel resolution and thus expect smaller location error in the mmWave and THz transmission range.
\item \textit{Latency reduction}: Since the location information is derived from the captured image, the complicated handshaking operations for the beam refinement process are unnecessary (see Fig.~\ref{fig:vision_aided_beam}), which means that the transmission latency is replaced by the DNN processing latency.
By employing dedicated AI-processors designed with a few nano-scale CMOS technology, we expect that the DNN processing latency can be further reduced in the near future.
Considering that the minimum latency of 5G-BM is $20\,\text{ms}$, CVBM can be an appealing option to support the ultra-reliable and low latency communications (URLLC) in 6G.
\item \textit{Energy saving}: In 5G-BM, the BS and mobile have to transmit multiple training beams to establish the beam alignment, consuming considerable energy just for the beam direction identification (e.g., around $20\,\text{W}$ for the mmWave CSI-RS beam transmission).
Since the beam refinement stage is replaced by the DNN processing, CVBM has a good potential to save the energy.
Energy saving gain might be higher in the THz communications since the THz transmission requires higher power to compensate for the path loss and complicated beam sweeping mechanism to generate the narrow pencil-type beam.
\end{itemize}

\section{Experiments and Discussions}

\subsection{Dataset and Experiment Setup}
\begin{figure*}[t]
	\centering
	\includegraphics[width=1\columnwidth]{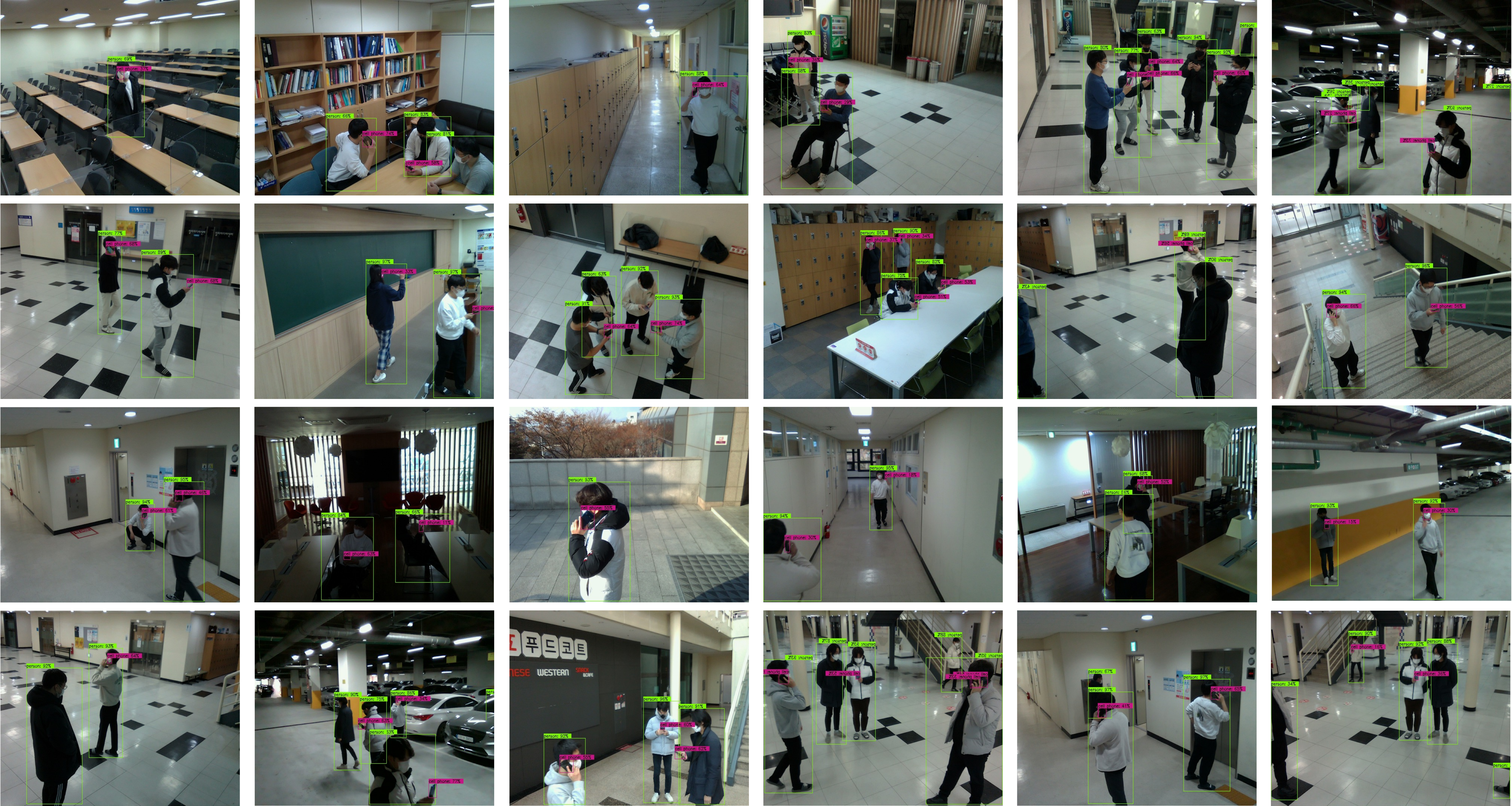}
	\vspace{-0.3cm}
	\caption{Collected data samples and their object detection results. Sample images are taken from various environments. Green and red bounding boxes indicate the person and the cell phone, respectively.}
	\label{fig:vision_dataset}
\end{figure*}

Since the main operation of the CVBM is the DL-based object detection, we should have a dataset to evaluate the effectiveness of the proposed approach.
To this end, we generated VOBEM dataset using the Intel RealSense L515 RGB-d camera.
The dataset consists of $135$ pairs of RGB and depth images acquired from 21 distinct environments including classroom, library, hallway, stair, and basement.
In each image, up to 6 people appear, each holding a mobile phone within a maximum distance of $8\,$m.
We manually annotate labels including class (human or a mobile phone), distance, angle, and bounding box on each image.
In Fig.~\ref{fig:vision_dataset}, we illustrate the snapshots of VOBEM datasets (see http://islab.snu.ac.kr/upload/vwdata.zip).

As an object detector, we use EfficientDet~\cite{efficientdet}, the state-of-the-art object detector outperforming the conventional ones such as Faster R-CNN~\cite{rcnn} and YOLO~\cite{yolo}.
Specifically, we use the model pre-trained on the MS-COCO 2017 dataset~\cite{mscoco} consisting of 80 classes of objects and 118,000 training images.
Using this DNN model, we perform the two-stage object detection to capture the large-scale object (person) and small-scale object (mobile) sequentially.

In the beam management simulations, we consider the scenario where the BS equipped with $N_{t}=64$ transmit antennas serves the mobile equipped with $N_{r}=4$ receiving antennas.
The location of the mobile is uniformly distributed in a square area of $20\times 20\,\text{m}^{2}$.
The carrier frequency of THz LoS channel model is $f_{c}=0.1\,\text{THz}$ with $1\,\text{GHz}$ bandwidth and the indoor path loss model specified in 3GPP TR 38.901 Rel. 16 is used.
For comparison, we use 5G-BM with $8$-bit DFT-based beam codebook with oversampling ratio being $4$.

\subsection{Experiment Results}
\begin{table*}[t]
    \centering
    \caption{Object localization performance using EfficientDet-D4 object detection model. Results are evaluated on the VOBEM dataset. Latency and power consumption are estimated based on the Qualcomm Snapdragon 888.}
    \vspace{-0.2cm}
    \resizebox{0.98\columnwidth}{!}{
    \begin{tabular}{c|cccc|cc|cc}
        \hline
        &
        \multicolumn{2}{c}{Human} & \multicolumn{2}{c}{Cell phone} & \multicolumn{2}{|c}{Localization error} & \multicolumn{2}{|c}{Resource usage} \\
          & Precision (\%) & Recall (\%) & Precision (\%) & Recall (\%) & Distance (cm) & Angle (deg.) & Latency (ms) & Power (W)   \\
         \hline
        CVBM & 94.96    & 96.08    & 96.68    & 90.67    & 3.74    & 0.23    & 15.8    & 10    \\
         5G-BM & - & - & - & - & 128.5 & 7.9 & 30 & 20 \\
         \hline
        \end{tabular}}
    \label{tab:vision_result}
\end{table*}

\begin{figure}
	\centering
	\subfloat[]{\includegraphics[width=.5\columnwidth]{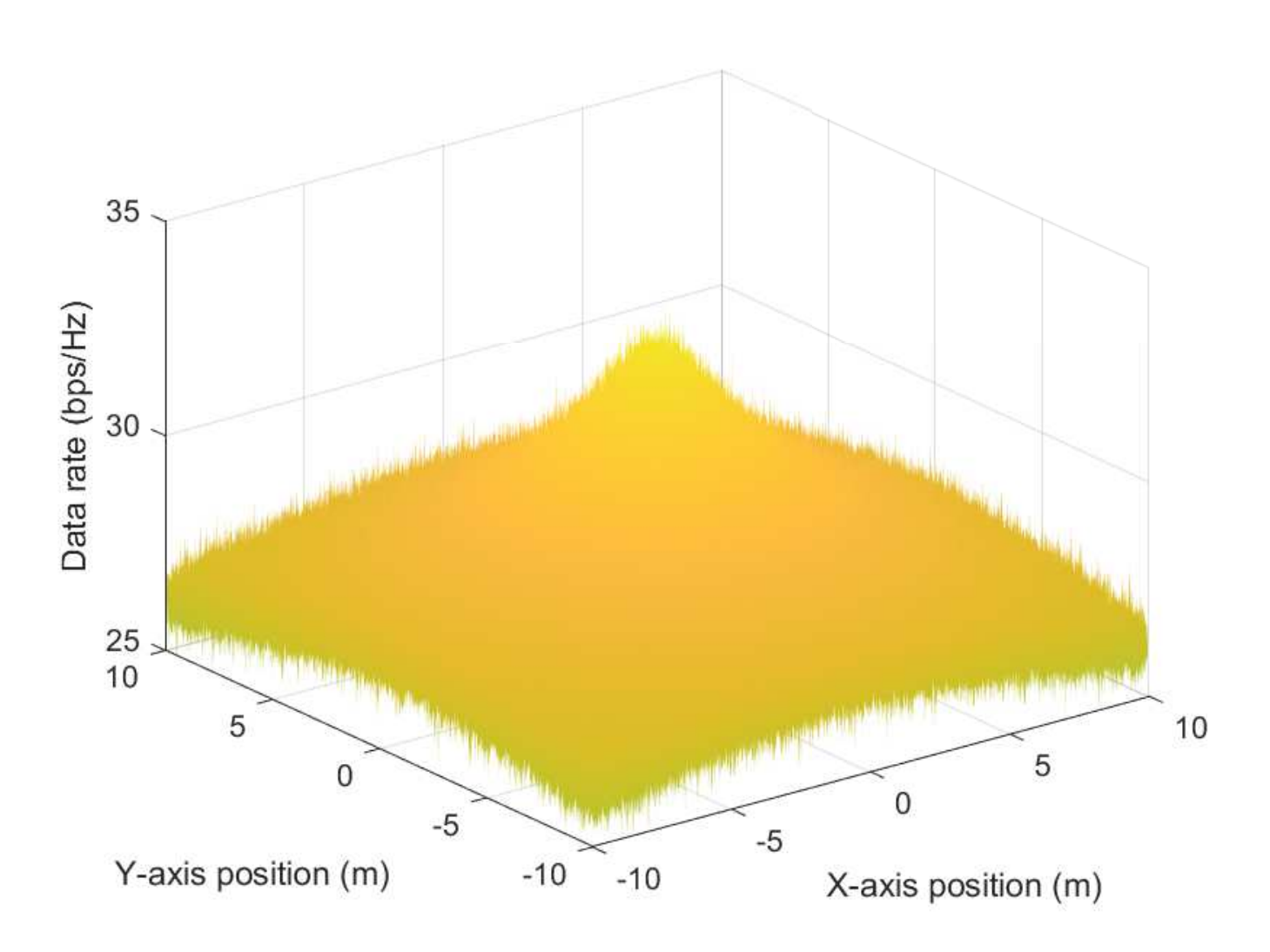}}
	\hfill
	\subfloat[]{\includegraphics[width=.5\columnwidth]{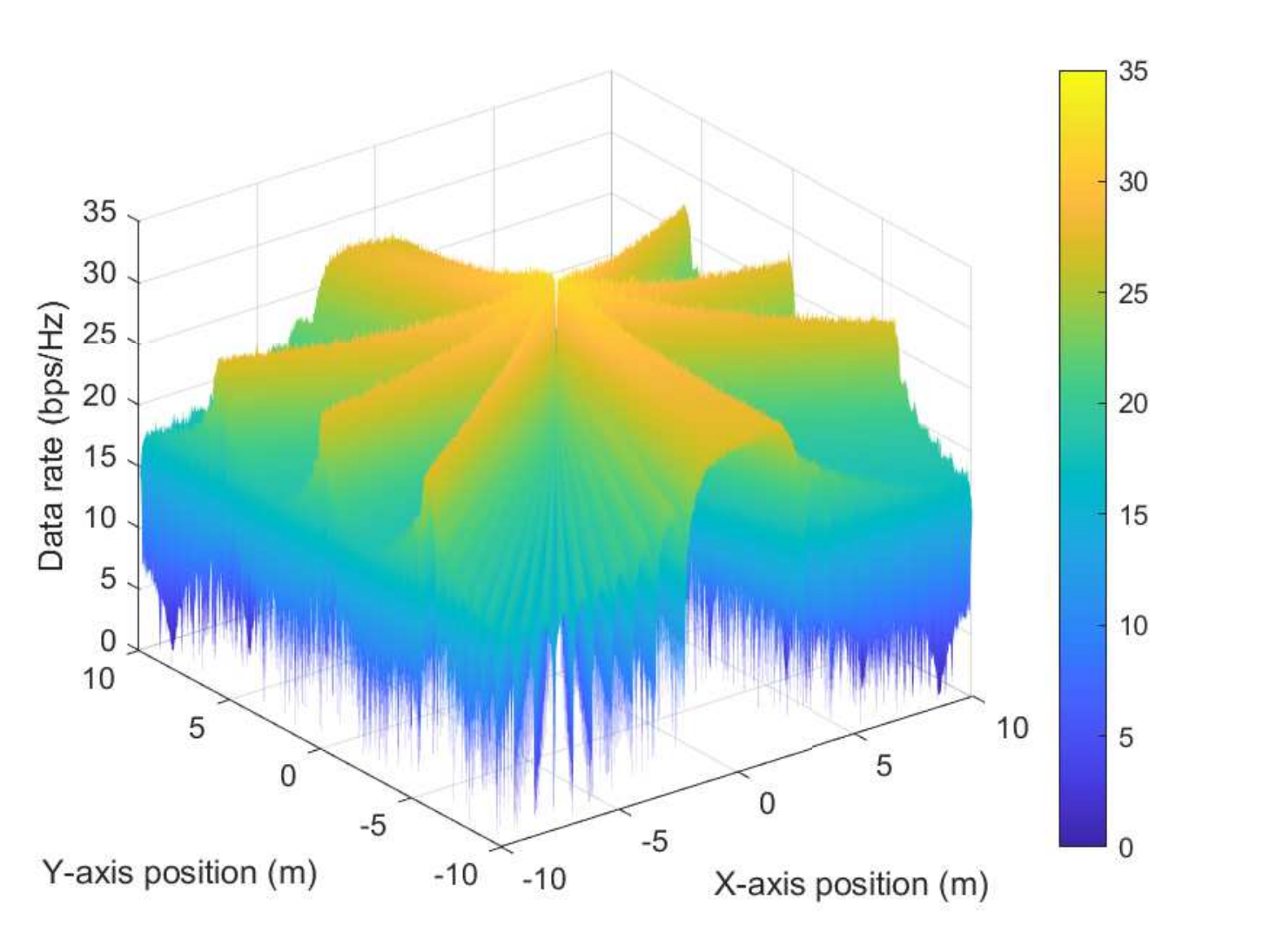}}
	\hfill
	\caption{(a) Illustration of data rate at each user position $(x,y)$ when the BS located at $(0,0)$ uses the vision-aided beam management technique. (b) Illustration of data rate at each user position $(x,y)$ when the BS located at $(0,0)$ uses the conventional beam management technique.}
	\label{fig:simul_results}
\end{figure}

In Table~\ref{tab:vision_result}, we summarize the object detection performance in terms of precision and recall.
The recall is the percentage of the detected objects among all target objects and the precision is the percentage of the correctly detected objects among total detected objects.
As shown in Table~\ref{tab:vision_result}, the DL-based object detector achieves more than 94\% of precision and recall on average.
Note, in the special corner case where the object detection fails (e.g., the case where the color of mobile phone and the background color are similar), the conventional beam management is used to serve the mobile (we discuss this issue in the next section).

In Table~\ref{tab:vision_result}, we compare 5G-BM and CVBM in terms of localization error, latency, and power consumption.
We observe that CVBM achieves more than $97\%$ improvement in the localization performance over 5G-BM.
As mentioned, a mismatch between the pre-defined beam direction and the real direction is unavoidable in the 5G-BM due to the codebook quantization but no such behavior occurs in CVBM since the beam is directly generated from the 3D localization information.
For example, in 5G-BM, the worst case beam direction error is around $7.9$ degrees but that of CVBM is around $0.2$ degrees.

We also see that the latency and the power consumption of the proposed scheme are much lower than those of the conventional scheme.
Since the location of a mobile is directly inferred from the captured image, complicated beam refinement and feedback operations are unnecessary, resulting in a reduction of the latency and transmit power. 
For instance, in the 5G NR, up to $4$ CSI-RS beams are transmitted with $10\,\text{ms}$ interval for the beam refinement and the corresponding latency is around $30\,\text{ms}$ but the inference time of CVBM is around $15.8\,$ms.

Based on the localization result in Table~\ref{tab:vision_result}, we evaluate the data rate of two beam management strategies as a function of user location.
In Fig.~\ref{fig:simul_results}, we depict the achievable rate of a mobile using the 3D color map.
We see that CVBM outperforms 5G-BM in all regions under test and in particular provides a seamless throughput for the whole service area.
Due to the finite resolution of the codebook, 5G-BM shows reliable performance only for the region where the mobile is well aligned with the beam direction.

\section{Conclusions and Future Directions}
In this article, we presented the CVBM as a new paradigm for the 6G wireless communication systems.
The main wisdom behind the proposed strategy is to replace the RF transmission-based beam control process by the powerful and proven AI processing.
From the numerical evaluations using the newly generated dataset called VOBEM, we demonstrated the effectiveness of CVBM in terms of data rate, latency, and energy consumption.
While we took the first step toward the AI-aided beam management, since the devil is in the detail, there are many important questions remaining to be answered.
We list here some of future research directions.

\begin{itemize}
\item CVBM dedicated DL model design:
Most of the DL-based object detectors to date focus on the identification of hundreds of objects (e.g., 80 classes in EfficientDet).
Since the quality of CVBM depends strongly on the object detection performance, it is of great importance to come up with a DL model suited for a small number of mobile objects.
Also, one can exploit the DL-based image masking, a technique to erase the undesired part of an image~\cite{sangtae_masking}.
Another approach worth investigation is the DL-based image super resolution (ISR) technique to convert the low-resolution image into the high-resolution one.
After the proper ISR processing, device identification and localization performance can be improved since the resolution of a mobile device occupying small pixels can be greatly enhanced.

\item CV-aided blockage detection and mobility management:
In our evaluations, we observe that the DL-based object detector does not perform well when the object is not on sight (e.g., the case where the mobile is under the desk).
When compared to 5G-BM requiring a long blockage detection latency, we believe that CVBM will readily handle the problem since the blockage detection can be quickly performed by checking the existence of the target mobile in an image.
Also, when the UE moves and thus the LoS blockage occurs, the BS can quickly launch the handover process using CVBM to switch the control of a mobile to the neighboring BS connected to the UE via LoS link.

\item Multi-modal sensing and acquisition:
In the 6G era, we expect that the network densification level will increase sharply so that the areal density of BSs will be comparable to, or even surpass, the density of mobile devices. 
In this ultra-dense network (UDN), the most of mobiles would be on sight of BS.
In the corner cases where the mobile is visually blocked, one can exploit the multi-modal information obtained from different types of devices such as ultrasonic sensor, thermographic camera, and infrared camera.
Also, a hybrid structure of CVBM and conventional RF transmission-based beam management can be used to identify the mobile objects in the blind spots.
For instance, CVBM can be selectively applied to the scenarios where the LoS link is available, just as the current non-standalone (NSA) architecture jointly uses the LTE and NR network to cover the case when the BS fails to establish the beam connection.

\item Vision processor design for fast and efficient CVBM:
In this article, we performed the proof-of-concept (PoC) evaluations based on the latency and energy consumption of the latest AI-focused SoC (i.e., \textit{Qualcomm Snapdragon 888}).
In order to verify the real-world latency and power consumption performances and also support a variety of emerging services in 6G, we should have a dedicated AI processor.
Low-power and low-latency AI processor will extend the application of CVBM to the unmanned aerial vehicles (UAVs) and IoT sensors.
\end{itemize}


\section*{Acknowledgements}
This work was supported by the National Research Foundation of Korea (NRF) grant funded by the Korea government (MSIT) (2022R1A5A1027646), the NRF grant funded by the Korea government (MSIT) (2020R1A2C2102198), and the MSIT (Ministry of Science and ICT), Korea, under the ITRC (Information Technology Research Center) support program (IITP-2022-2017-0-01637) supervised by the IITP (Institute for Information \& communications Technology Promotion).

\bibliographystyle{IEEEtran}

\end{document}